# Transient and secular radioactive equilibrium revisited

Qinghui Zhang, Ph.D.,
Radiation Oncology Department, University of Nebraska Medical Center, Omaha 68198

Departments of Medical Physics, Memorial Sloan-Kettering Cancer Center,
New York, NY 10065

Chandra Burman, Ph.D.,
Departments of Medical Physics, Memorial Sloan-Kettering Cancer Center,
New York, NY 10065

Howard Amols, Ph.D.
Department of Radiation Oncology, Columbia University Medical Center,
New York, Ny 10032

Departments of Medical Physics, Memorial Sloan-Kettering Cancer Center,
New York, NY 10065

**Abstract:** *The two definitions of radioactive equilibrium are revisited in this paper. The terms "activity equilibrium" and "effective life equilibrium" are proposed to take the place of currently used terms "transient equilibrium" and "secular equilibrium". The proposed new definitions have the advantage of providing a clearer physics meaning. Besides the well known instant activity equilibrium, another class of exact effective life-time equilibrium is also discussed in this letter.*

Radioactive decay is one of the most important phenomena in the atomic and nuclear physics [1]. Since the paper [7] of Amols and Bardash[7], several publications have discussed the confusions of the definitions of transient and secular radioactive equilibrium [1-11]. It has also been pointed out [6, 9, 10] that an exact equilibrium only exists for an instant in time; however, some kind of approximate equilibrium may exist (called pseudo-radioactive equilibrium in [9]). The purposes of this letter are (1) to clarify the confusions of the definitions of transient and secular radioactive equilibrium (2) to give physics explanations for these definitions. To do this, in this letter, we examine the definitions of radioactive equilibrium and propose new definitions that have a clearer physics meaning. We also show that there is another class of exact equilibrium besides the well known instant equilibrium.

For a decay process $a \to b \to c$, the numbers of parent $N_a(t)$ and daughter $N_b(t)$ nuclei change according to the following equations:

$$\frac{dN_a(t)}{dt} = -\lambda_a N_a(t) \quad , \tag{1}$$

and

$$\frac{dN_b(t)}{dt} = \lambda_a N_a(t) - \lambda_b N_b(t). \tag{2}$$

Here $\lambda_a$ and $\lambda_b$ are the decay constants of the parent and daughter respectively. The solutions of Eqs (1, 2) are

$$N_a(t) = N_a(0)e^{-\lambda_a t} \qquad (3)$$

and

$$N_b(t) = \begin{cases} N_b(0)e^{-\lambda_b t} + \dfrac{\lambda_a}{\lambda_b - \lambda_a} N_a(0)(e^{-\lambda_a t} - e^{-\lambda_b t}) & \lambda_a \neq \lambda_b \\ N_b(0)e^{-\lambda t} + N_a(0)t\lambda e^{-\lambda t} & \lambda_a = \lambda_b = \lambda \end{cases} \qquad (4)$$

Because $\lambda_a = \lambda_b$ seldom occurs in the natural world, we will ignore this situation in this letter. Eq. (1) and (2) can be rewritten as

$$\frac{d \ln N_a(t)}{dt} = -\lambda_a, \qquad (5)$$

$$\frac{d \ln N_b(t)}{dt} = \lambda_a \frac{N_a(t)}{N_b(t)} - \lambda_b. \qquad (6)$$

The right hand sides (RHS) of Eq. (1) and Eq. (2) always approximate to zero when time goes to infinity; however, this is not always the case for the RHS of Eq. (5) and Eq. (6). We point out that Eq. (5) and Eq. (6) remain valid if we change the numbers of parent and daughter nuclei to activity, given by $A_a(t) = \lambda_a N_a(t)$ and $A_b(t) = \lambda_b N_b(t)$ respectively. Then we have

$$\frac{d \ln A_a(t)}{dt} = -\lambda_a, \qquad (7)$$

$$\frac{d \ln A_b(t)}{dt} = \lambda_a \frac{N_a(t)}{N_b(t)} - \lambda_b. \qquad (8)$$

In radioactive decay, there are two different definitions of equilibrium [1-5]:

(1) Assuming the RHS of Eq. (2) is zero, i.e., the number of daughter nuclei or daughter activity is a constant [2, 3], one obtains $\lambda_a N_a(t_e) = \lambda_b N_b(t_e)$. It is found that at time

$$t_e = \frac{1}{\lambda_a - \lambda_b} \ln\left[\frac{\lambda_a^2 N_a(0)}{\lambda_b((\lambda_a - \lambda_b)N_b(0) + \lambda_a N_a(0))}\right] \quad (9a)$$

or (when $N_b(0) = 0$)

$$t_e = \frac{\ln(\frac{\lambda_a}{\lambda_b})}{\lambda_a - \lambda_b} \quad (9b)$$

the parent and daughter have the same activity, called transient equilibrium for its short duration and exists only when $\lambda_a > \lambda_b$ as observed from Eq. (9). If $\lambda_a \approx 0$ and $\lambda_a < \lambda_b$, then $\lambda_a N_a(t) \approx \lambda_b N_b(t)$ for very large t; that is daughter activity is slightly bigger than its parent's. This is called secular equilibrium because of its long duration (This is a pseudo-radioactive equilibrium according to [9]).

(2) When the ratio of the activities of daughter to parent is constant with time, daughter and parent are said to be in equilibrium [1, 4, 5]. Defining

$$y = \frac{A_b(t)}{A_a(t)} = \frac{\lambda_b N_b(t)}{\lambda_a N_a(t)} = \frac{\lambda_b N_b(0)}{\lambda_a N_a(0)} e^{(\lambda_a - \lambda_b)t} + \frac{\lambda_b}{\lambda_b - \lambda_a}(1 - e^{(\lambda_a - \lambda_b)t}), \quad (10)$$

one obtains $\frac{dy}{dt} = \left[\frac{\lambda_b N_b(0)}{\lambda_a N_a(0)}(\lambda_a - \lambda_b) + \lambda_b\right] e^{(\lambda_a - \lambda_b)t}$ by using Eqs (3) and (4). It is

clear that for $t \gg \frac{1}{\lambda_b - \lambda_a}$ and $\lambda_a < \lambda_b$, $\frac{dy}{dt} \approx 0$ and $y = \frac{\lambda_b}{\lambda_b - \lambda_a}$; thus an approximate equilibrium is obtained. The case where $\frac{\lambda_a}{\lambda_b}$ is very small (i.e., parent lifetime is long relative to daughter) and y approaches unity is called secular equilibrium. The case where $\frac{\lambda_a}{\lambda_b}$ is large and y is not unity is called transient equilibrium because of the relatively short lifetime of the parent to daughter.

In the following, we refer to the above two definitions as **Definition I** (i.e., daughter activity is constant) and **Definition II** (i.e., ratio of daughter activity to parent activity is constant). Because the physics meaning of Definition II is not clear, in the following, we will concentrate on the **Definition II** and try to give it a physics explanation. Figure 1 shows the ratio *y(t)* of daughter activity to parent activity vs. time for three different cases. The solid line corresponds to the case of secular equilibrium, the dot-dashed line to transient equilibrium and the dashed line to no equilibrium. According to **Definition II**, we notice that even when the parent is the same nuclei, it could be in secular, transient, or no equilibrium depending on the lifetime of different daughters. The plot of the two different activities (not the ratio) as a function of time, widely used in current textbooks, is more appropriate in the context of **Definition I.**

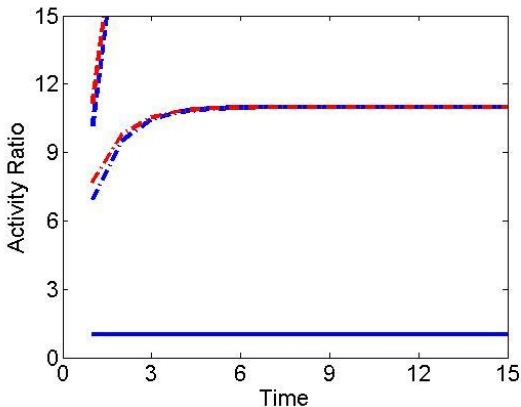

Fig.1: The ratio of daughter activity to parent activity vs. time. The solid line corresponds to the case $\lambda_a = 10$ and $\lambda_b = 100000$ with $\frac{N_b(0)}{N_a(0)} = 0.2$ and $N_b(0) = 0$ (both fall on almost

the same line in the figure), the dot-dashed line to $\lambda_a = 10$ and $\lambda_b = 11$ with $\frac{N_b(0)}{N_a(0)} = 0.2$ (red) and $N_b(0) = 0$ (blue), and the dashed line to $\lambda_a = 10$ and $\lambda_b = 9.999999$ with $\frac{N_b(0)}{N_a(0)} = 0.2$ (red) and $N_b(0) = 0$ (blue). Here the time unit is arbitrary.

Current textbooks describe two types of equilibrium (**Definitions I** and **II** given above), where each type of equilibrium is further separated into transient and secular equilibrium according to its duration. This leads to confusion if one simply uses transient and secular equilibrium without mentioning which type of equilibrium [6-11]. Transient and secular equilibrium have been used for a long time without confusion until the emergence of **Definition I**. In physics, an equilibrium definition normally is named by its unchanged observation and not by its duration of time. Because of the confusion of transient equilibrium and secular equilibrium, we propose to give a name to each type of equilibrium. For **Definition I**, since the activities of daughter is constant (i.e, activities of the parent and daughter are the same from Eq. (2)), we refer to it as "activity equilibrium". The physics meaning of **Definition II** will become apparent below.

Define a new variable $z(t) = \ln y(t) = \ln A_b - \ln A_a$ and substituting Eqs. (7, 8), we obtain $\frac{dz(t)}{dt} = \frac{d \ln A_b}{dt} - \frac{d \ln A_a}{dt} = \lambda_a \frac{N_a(t)}{N_b(t)} - \lambda_b + \lambda_a$. Notice that the RHS of Eq. (7) tells us the lifetime of the parent (because mean lifetime is $\frac{1}{\lambda_a}$). We define $\frac{d \ln(activity)}{dt}$ as the effective life time, (from Eq. (7)) which is exactly equal to the mean lifetime for the parent and (from Eq. (8)) is only approximately true for the daughter nuclei. However as shown below, the effective lifetime equals the daughter after a long time. Thus $\frac{dz(t)}{dt} = 0$ reflects the equivalence between the effective lifetimes of the parent and daughter. Therefore, definition two can be called the effective lifetime equilibrium. The effective decay parameter (which is a constant) of the parent is $\lambda_a$ and the effective decay parameter (which is not a constant) of the daughter is (when $\frac{N_b(0)}{N_a(0)} \neq \frac{\lambda_a}{\lambda_b - \lambda_a}$)

$$\lambda_{b,eff} = -\lambda_a \frac{N_a(t)}{N_b(t)} + \lambda_b = -\lambda_a \frac{1}{\left(\frac{N_b(0)}{N_a(0)} - \frac{\lambda_a}{\lambda_b - \lambda_a}\right) e^{(\lambda_a - \lambda_b)t} + \frac{\lambda_a}{\lambda_b - \lambda_a}} + \lambda_b = \begin{cases} \lambda_b & \text{if } \lambda_a > \lambda_b \; t \to \infty \\ \lambda_a & \text{if } \lambda_a < \lambda_b \; t \to \infty \end{cases}$$

(11a)

or (when $N_b(0) = 0$)

$$\lambda_{b,eff} = -\lambda_a \frac{N_a(t)}{N_b(t)} + \lambda_b = -\frac{\lambda_b - \lambda_a}{1 - e^{(\lambda_a - \lambda_b)t}} + \lambda_b = \begin{cases} \lambda_b & \text{if } \lambda_a > \lambda_b \; t \to \infty \\ \lambda_a & \text{if } \lambda_a < \lambda_b \; t \to \infty \end{cases} \quad (11b)$$

We define $D(t) = \frac{1}{\lambda_{b,eff}} - \frac{1}{\lambda_{a,eff}}$ or $R(t) = \frac{\lambda_{b,eff}}{\lambda_{a,eff}}$ to reflect the differences between the effective life time of the daughter and parent. In Fig.2, D vs. time is shown. One observes that when $\lambda_a < \lambda_b$ and t goes to infinity, D approaches zero because the daughter has an effective life time similar to the parent (solid line and dot dashed line). It is interesting thing to notice that although the activity ratio is quite different (solid and dot dashed lines in Fig.1), their effective life time at large times is the same. When $\lambda_a > \lambda_b$ and time goes to infinity, D is the difference between the daughter and parent (dashed line). In Fig.1, when $\lambda_b > \lambda_a$, we obtain a different ratio of parent activity to daughter activity for different $\lambda_b$ when time goes to infinity. On the other hand, the difference in parent and daughter effective lifetimes remains the same for different $\lambda_b$ in Fig.2. Therefore, Fig.2 clearly tells us that the effective lifetime is an appropriate name for the definition II.

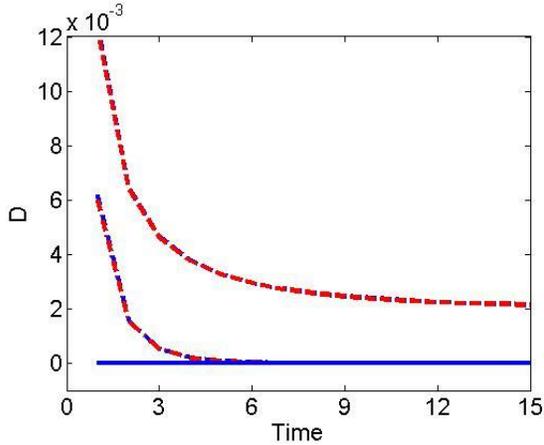

Fig.2: The difference between effective lifetime of daughter and parent, $D = \dfrac{1}{\lambda_{b,eff}} - \dfrac{1}{\lambda_a}$, vs. time. The solid line corresponds to cases $\lambda_a = 10$ and $\lambda_b = 100000$ with $\dfrac{N_b(0)}{N_a(0)} = 0.2$ and $N_b(0) = 0$ (both fall on almost the same line in the figure), the dot-dashed line to $\lambda_a = 10$ and $\lambda_b = 11$ with $\dfrac{N_b(0)}{N_a(0)} = 0.2$ (red) and $N_b(0) = 0$ (blue) and the dashed line to $\lambda_a = 10$ and $\lambda_b = 9.8$ with $\dfrac{N_b(0)}{N_a(0)} = 0.2$ (red) and $N_b(0) = 0$ (blue). Here the time unit is arbitrary.

From Eq. (11a), we notice an interesting phenomenon. If

$$\dfrac{N_b(0)}{N_a(0)} = \dfrac{\lambda_a}{\lambda_b - \lambda_a} \; , \qquad (12)$$

Then

$$\lambda_{b,eff} = \lambda_a \qquad (13)$$

for all time. It is known from Eq. (11a) that if $\lambda_b > \lambda_a$ then $\lambda_{b,eff} = \lambda_a$ when t is large. However Eq. (12) tells us that for this special value of initial daughter number, $\lambda_{b,eff} = \lambda_a$ for all time. This can be seen clearly from Fig. 4. One can also prove that the ratio of daughter to parent activity $y = \dfrac{\lambda_b N_b(t)}{\lambda_a N_a(t)} = \dfrac{\lambda_a}{\lambda_b - \lambda_a}$ for all time. Therefore, besides the well known exact equilibrium at an instant in time (see Eq. (9)), this is another exact equilibrium posting to definition II, referred to here as exact effective lifetime equilibrium. This exact equilibrium holds true for all times after when $\dfrac{N_b(0)}{N_a(0)} = \dfrac{\lambda_a}{\lambda_b - \lambda_a}$.

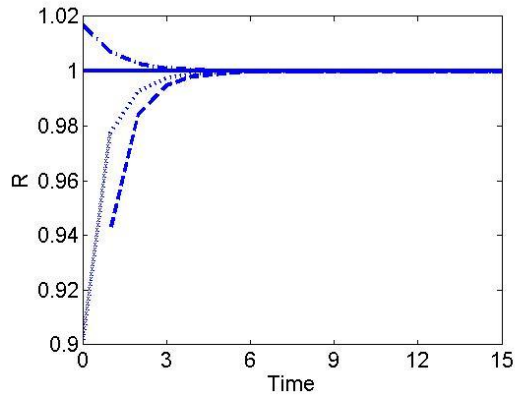

Fig.4: The ratio of the daughter effective decay parameter to the parent decay parameter vs. time, $R = \dfrac{\lambda_{b,eff}}{\lambda_a}$. Here $\lambda_a = 10$ and $\lambda_b = 11$. Here dashed line, dotted line, solid line and dot-dashed line corresponds to $\dfrac{N_b(0)}{N_a(0)} = 0.0$, $\dfrac{N_b(0)}{N_a(0)} = 5.0$, $\dfrac{N_b(0)}{N_a(0)} = \dfrac{\lambda_a}{\lambda_b - \lambda_a} = 10$, and $\dfrac{N_b(0)}{N_a(0)} = 12$ respectively. It is interesting to notice that when $\dfrac{N_b(0)}{N_a(0)} = \dfrac{\lambda_a}{\lambda_b - \lambda_a} = 10$, R is one from the beginning of the decay process.

Given these definitions, we can classify the two different types of transient equilibrium and secular equilibrium. By coincidence, secular equilibrium of **Definition I** (activity equilibrium) and secular equilibrium of **Definition II** (effective lifetime equilibrium) is similar. This can be understood by noting that if the daughter activity is the same as the parent activity, then the daughter decay behavior is similar to its parent, and the daughter's effective life time must be the same as that of the parent. We make the following observations:

(1): As stated clearly in Ref. [6], this letter is only of academic interest. Because the number of decay processes is limited, we can simply define transient and secular equilibrium for each process. On the other hand, physics is also a logical science. The definition of physical concepts should be very strict. Therefore, we think the new definition given here would be more appropriate in textbook descriptions.

(2): Even with these new definitions, we do not suggest dividing each equilibrium into two classes, namely, transient and secular; since this classification can lead to confusion [7]. Instead we propose referring to first class of them as activity equilibrium and effective life-time equilibrium, respectively. For secular equilibrium of both cases, we can call them "activity equilibrium and effective lifetime equilibrium" because activity equilibrium and effective lifetime equilibrium occur at the same time.

(3): We emphasize that the equilibrium here is an approximate one. There is only exact equilibrium at one time instant in present textbook. In this letter, we have noted another class of exact equilibrium.

In conclusion: Two definitions of radioactive equilibrium are proposed in this letter, which we refer to as activity equilibrium and effective life-time equilibrium, respectively. Given these proposed definitions, we believe that the confusion related to transient and secular equilibrium no long exist. We further observe that in addition to the well known activity equilibrium at one time instant, there is also another exact effective life-time equilibrium.

Acknowledgement: The authors like to thank Dr. W. R. Hendee and Dr. D. R. Bednarek for helpful communications. The authors also like to thank Dr. G. S. Mageras for his helpful discussions and help in writing.